# Learning strategies for global games with delayed payoffs

Wan Ahmad Tajuddin WAN ABDULLAH

Physics Department, Universiti Malaya, 50603 Kuala Lumpur, Malaysia

## Abstract

Economic ensembles can be modeled as networks of interacting agents whose behaviors are described in terms of game theory. The evolutionary paradigm has been applied to two-person games to discover strategies in this context. Subsequently, many-player games, and specifically global games (where payoffs depend collectively on all the rest of the players) as with the minimal game, have been studied. The minority game is attractive because it has intuitive similarities to e.g. securing niche businesses. We enhance this intuitive similarity by extending the game through the introduction of delayed payoffs. Payoffs depend on the values of future moves; we reward choices which later on become popular.

We study agents' moves in such global game with delayed payoff. Instead of an evolutionary approach we allow learning in the agents. We study strategies that may emerge through learning in agents in such games.

## 1 INTRODUCTION

From a physics viewpoint, dynamical systems can be understood to behave asymptotically in three ways described by point attractors, cyclic attractors and strange attractors. Economic ensembles, having large numbers of degrees of freedom, and probably nonlinear interactions, are dynamical systems. Classical economics can be associated with point attractors, while phenomena like business and economic cycles can perhaps be seen as manifestations of cyclic attractors. Of current interest is the regime of strange attractors, where complex dynamics can either produce virtually random trajectories or complex trajectories with fractal characteristics.

The study of the complex dynamics of economic ensembles consisting of interacting agents whose behaviors are described in terms of games, has been initiated through the application of the evolutionary paradigm to the two-person prisoners' dilemma game [1]. Economic ensembles can be modeled as networks of many interacting agents playing games, suggesting the use of many-player games. There have been studies [5] of many-player prisoners' dilemma, but prisoners' dilemma being inherently two-player, each player here interacts only with his immediate neighbors. From the physics point of view, elements in the network interact only first-order and/or locally. One interesting many-player game studied is the minority game [2] where payoffs depend on the actions of all other players, thus yielding a network with higher-order or global interactions. The minority game is attractive because it has intuitive similarities to e.g. securing niche businesses.

In this paper, we introduce a new class of games, in which in addition to having global interactions, payoffs are delayed in the sense that they depend on the values of the players' future moves. In particular, subscribing to some economic intuition, we design payoffs in such a way that we reward (near-) minority choices which later on become popular. We study agents' or players' long-time moves in such a game. We use learning agents instead of evolutionary ones and we vary the amount of information available from which the agents can learn. We also explore the effect of culture, understood as direct transmission of knowledge between agents.

## 2 THE MODEL

$N$ players or agents $k$ each chooses one of $n$ options $i$. If $i_k(t)$ is $k$'s choice at timestep $t$, then payoffs $p_i$ for moves $i$ at $t$ is taken to be

$$p_i(t) = \sum_k \delta(i, i_k(t+1)) - \sum_k \delta(i, i_k(t)) \qquad (1)$$

where the Kronecker delta $\delta(x,y) = 1$ when $x = y$ and 0 otherwise.

Agents learn using a neural network without hidden units, by employing a variant of the back-propagation algorithm [6], or equivalently, Hebbian learning with nonlinearity. For an input $u$ contributing to the output $i_k(t)$, the synaptic strength $T^k_{ui}$ between them changes by

$$\Delta T^k_{ui} = \eta (p_i - h^k_i) f'(h^k_i) + \chi \iota \qquad (2)$$

where $\eta$ is the learning rate, $h^k_i$ is the field for the neuron representing $i_k$,

$$h^k_i = \sum_u T^k_{ui} u, \qquad (3)$$

$f'(x)$ is the differential of the neuron transfer function (the nonlinearity), taken to be

$$f'(x) = 1/(x^2+1), \qquad (4)$$

and $\chi$ is a 'creativity' factor incorporating 'noise' or some 'non-zero temperature' effects via the random number $\iota$ picked from a uniform random distribution be-tween 0 and 1. Initial synaptic values are small random numbers equally likely to be positive or negative. Without learning ($\eta = 0$), option choices are random.

Agents decide on options through a competitive network [4] involving the final layer neurons; in particular, the agent $k$ chooses the option $I$ if $h^k_I = \max(\{h^k_i\}_i)$. We further defined the memory $M$ of agents as the number of timesteps back that they remember of payoffs and moves. Cases where $M = 0$ then are equivalent to random choices.

Inputs $u$ to the field for decision and to the learning may consist only of each respective agent's own previous moves ('self learning'). A more plausible scenario is where $u$ also include previous resulting payoffs ('learning with payoff'), and with information, there is also the case where in addition to these, the previous moves of other agents contribute to the decision-making and the learning ('learning with information'). A model of culture is also explored where there is a direct contribution to the 'knowledge' stored in an agent's synapses from those of the others:

$$T^k_{ui} := \varepsilon \left(\sum_{k' \neq k} T^{k'}_{ui}\right) + (1-\varepsilon)\, T^k_{ui} \tag{5}$$

with $\varepsilon$ being the 'culture factor' and this inclusion of culture is carried out after learning (with information) is done ('learning with culture).

## 3 SIMULATION AND RESULTS

We carried out studies on the model using computer simulations. Specifically, 10 agents choose between 5 options. The following parameters were used: initial randomness (maximum magnitude of initial random value for synapses) 0.1, creativity 0.1, culture factor 0.2 and learning rate 0.1. We studied (100 random trials) long-time behavior (after 1000 timesteps) of cases of no learning, self learning, learning with payoff, learning with information and learning with culture, each with zero memory, and memories of 1, 3 and 7. We investigated the mean (between agents) average (over timesteps) payoffs obtained by agents, which measures the average performance of the players, and the standard deviation (between agents) of the average payoffs, which gives a measure of variety of the performances.

For the cases of no learning, we obtain a well-defined cluster of means of average payoff at around -0.8 and standard deviations between about 0.02 to 0.08. This is shown in the figure below. Learning with zero memory also yields a similar output, as expected.

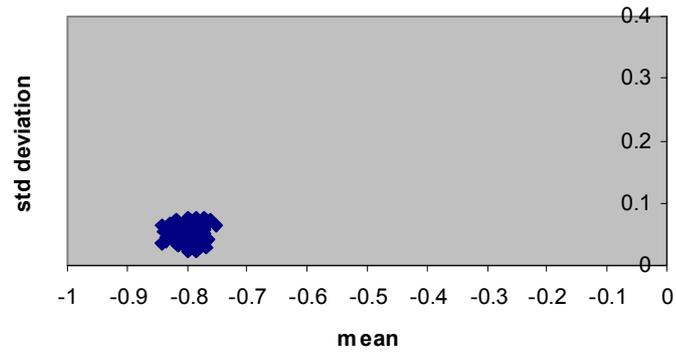

**Fig. 1.** Standard deviation vs mean of average cumulative payoffs for the case of no learning and memory 0

Self learning generally betters the average performances without significantly increasing variances. This is portrayed in the following graphs. An interesting thing to note is the better average performance when only immediately recent states are remembered.

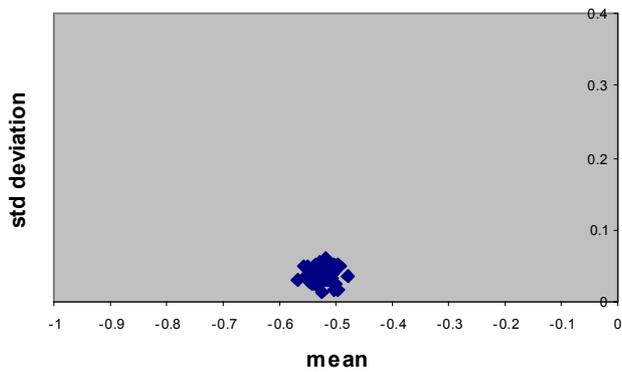

**Fig. 2.** Standard deviation vs mean of average cumulative payoffs for the case of self learning and memory 1

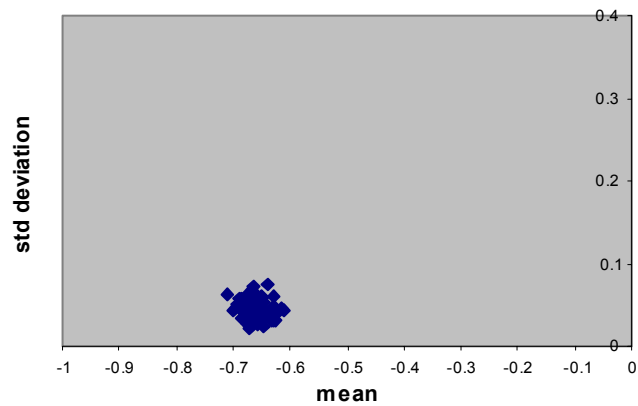

**Fig. 3.** Standard deviation vs mean of average cumulative payoffs for the case of self learning and memory 3

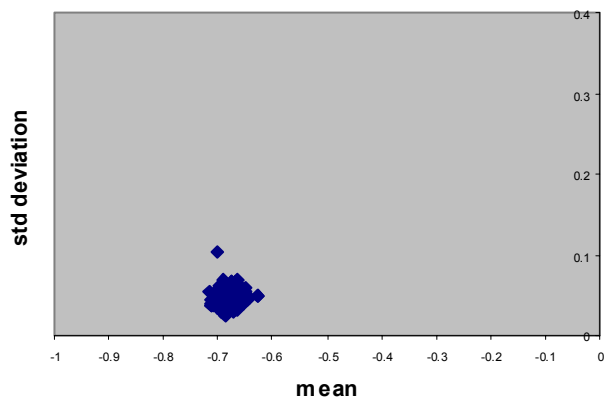

**Fig. 4.** Standard deviation vs mean of average cumulative payoffs for the case of self learning and memory 7

Given knowledge of resulting payoffs, agents should learn to perform better. This is depicted by higher means, as is evident in the following graphs. Increased standard deviations, reflecting emergence of variety, can also be seen, although with more memories, this seems to decrease.

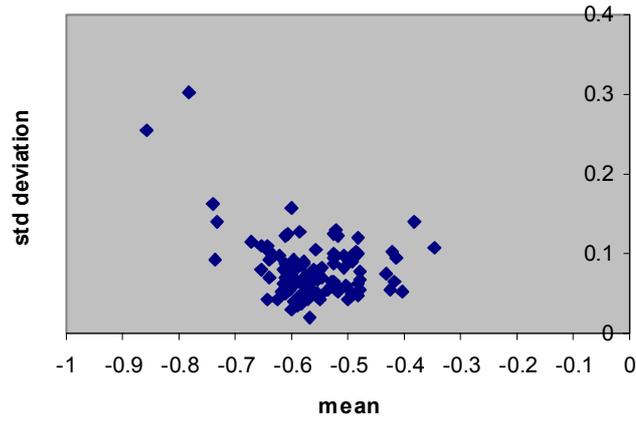

**Fig. 5.** Standard deviation vs mean of average cumulative payoffs for the case of learning with payoff and memory 1

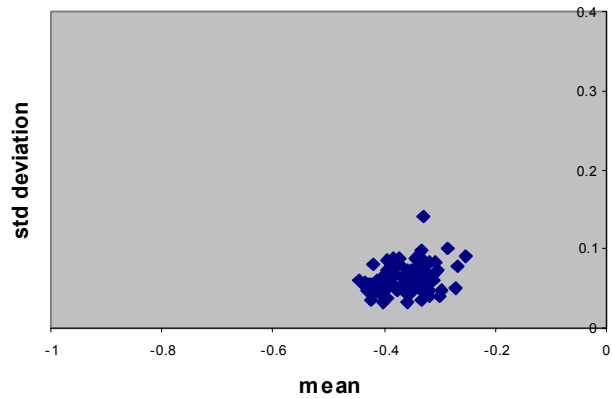

**Fig. 6.** Standard deviation vs mean of average cumulative payoffs for the case of learning with payoff and memory 7

Learning with information shows similar behavior to learning with payoff, both with means as well as standard deviations, almost quantitatively.

Learning with culture smears the means and makes them vary substantially, but the standard deviations remain more or less in the band defined by no learning. The results of learning with culture with $M$=3 is shown below. Note the different scale for the mean.

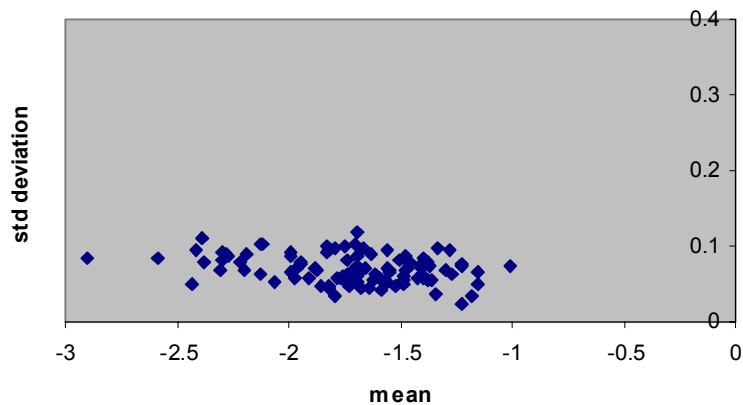

**Fig. 7.** Standard deviation vs mean of average cumulative payoffs for the case of learning with culture and memory 3

In terms of strategies learnt, one (which is easy to detect) is that of holding on to the same option. We find that some winning agents employ this strategy, though employing this strategy doesn't seem to guarantee winning all the time. The existence or not of less simplistic strategies requires further studies.

## 4 CONCLUSION AND DISCUSSION

We have introduced a new class of games, where payoffs are delayed. Our studies show how agents can learn to perform better in such games, even without outside information. Giving information on complete payoff vectors seem to be almost equivalent to also giving information on other agents' moves. Information as such increases agents' average performance as well as brings rise to increased variety. Heterogeneity in agents can perhaps be explained through such emergence. Culture as modeled here does not give rise to heterogeneity between agents, but does give rise to an array of heterogeneous overall performances. Because of culture agents behave similarly, but these common behaviors give rise to varied performances between trials.

Evolutionary prisoners' dilemma has brought about the surprising emergence of cooperation, while the minority game has shown the emergence of groups with extreme behaviors [3]. Perhaps the emergence of heterogeneity can be studied using the game proposed in this paper.